\documentstyle[12pt]{article}
\begin{document}
\thispagestyle{empty}
\begin{center}
\large{Classification of Two Dimensional $N=4$ Superconformal Symmetries}
\end{center}
\vspace*{1cm}
\begin{center}

\sc{Abbas Ali}

\it{ Physics Department,\\ Aligarh Muslim University, \\Aligarh 202002,
India}
\end{center}
\vspace*{2.5cm}
\begin{center}
\large{Abstract}
\end{center}

Classification of $N=4$ superconformal symmetries in two 
dimensions is re-examined. It is proposed that
apart from $SU(2)$ and $SU(2)\times SU(2)\times U(1)$ their Kac-Moody
symmetry can also be $SU(2)\times(U(1))^4$. These superconformal 
symmetries and corresponding algebras are named small, large and middle 
ones respectively. Operator product expansions for the middle algebra are
derived. Complete free field realizations of large and middle
superconformal symmetries are obtained.

\vspace*{5cm}

\newpage

Conformal symmetries in two space-time dimensions have been very
extensively studied owing to their applications both in string theory
and two dimensional statistical systems. The same is true
for their supersymmetric generalisations which have been classified
at least upto $N=3$ superconformal symmetries in the sense 
that we know the
corresponding superconformal graded Lie algebras, operator product
expansions (OPEs) and the free field realizations. Same, however, can not be
said for theories with $N=4$ superconformal symmetries.
This situation is remedied
in this note. This should also complete the classification of all
two dimensional linear superconformal theories because $N=4$ 
symmetries are the highest ones which can have non-vanishing central charges
without requiring operators of negative dimensions.

Classification of $N=4$ superconformal symmetries is very interesting
because, unlike the $N=1, 2$ and 3 cases, there exist three 
such symmetries. Easiest way to 
distinguish among them is to specify their Kac-Moody symmetries.
An $N=4$ superconformal symmetry can have either $SU(2)$ or
$SU(2)\times SU(2)\times U(1)$ or $SU(2)\times(U(1))^4$ Kac-Moody
subsymmetry. We shall refer
to corresponding superconformal symmetries (algebras) as small, large
and middle ones respectively. Out of these symmetries the small and large
ones are well known and well established. Our aim in this note is to
obtain minimal information required to put the middle symmtery on
equal footings with them. Redressal of this problem acquires certain 
urgency because of abundant occurrence of $N=4$ superconformal 
symmetries in the context of AdS-CFT correspondence 
\cite{mcena}-\cite{adscft4} related to three dimensional stringy black 
hole \cite{adscft3, ak0} backgrounds as well as D1/D5 system
\cite{hw, sw}.

This paper is organised as follows. First of all we very briefly
describe the small $N=4$ superconformal symmetry algebra and present
corresponding (anti-) commutation relations, OPEs
and a suitable free field realization. Next we do the same
for the large algebra. After that we list the (anti-) commutation 
relations of the middle algebra and derive corresponding OPEs.
A free field realization of this symmetry is then
obtained. This realization turns out to be incomplete.
We identify the source of this problem and then take 
appropriate steps to remove the discrepancy.
To do it we start by generalizing the most
commonly used free field realization of the large algebra. A complete
free field realization of the middle algebra is then obtained.
Finally conclusions about complete classification of $N=4$ superconformal 
symmetries in two dimensions and suggestions for further investigations 
are presented.

The small $N=4$ superconformal symmetry was discovered by Ademollo
et al \cite{adm}. It is generated by eight operators
consisting of the energy-momentum tensor $\tilde{T}(z)$, its four
superpartners $\tilde{G}^a(z), a\in \{0,i\}, i\in\{1, 2, 3\}$
and three $SU(2)$ Kac-Moody currents $\tilde{J}^i(z)$. Corresponding 
holomorphic anomalous dimensions are 2, 3/2 and 1 respectively. 
Modes of these operators are denoted by $\tilde{L}_m$, $\tilde{G}^a_r$
and $\tilde{J}^i_m$ respectively. These modes satisfy the following
small $N=4$ graded Lie superconformal algebra.
\begin{eqnarray}
\left[\tilde{L}_m, \tilde{L}_n\right]& = &
(m-n)\tilde{L}_{m+n}+\frac{\tilde{c}}{12}
(m^3-m)\delta_{{m+n},0},\nonumber\\
\left[\tilde{L}_m, \tilde{G}_r^a\right]& = &
\left(\frac{m}{2}-r\right)\tilde{G}^a_{m+r},
~~~\left[\tilde{L}_m, \tilde{J}^i_n\right]=-n\tilde{J}^i_{m+n},\nonumber\\
\{\tilde{G}^a_r, \tilde{G}^b_s\}& = &2\delta^{ab}\tilde{L}_{r+s}
-4(r-s)\alpha^{+i}_{ab}\tilde{J}^i_{r+s}+\frac{\tilde{c}}{3}
\left(r^2-\frac{1}{4}\right)\delta^{ab}\delta_{{r+s},0},\nonumber\\
\left[\tilde{J}^i_m, \tilde{G}^a_r\right]& = &
\alpha^{+i}_{ab}\tilde{G}^b_{m+r},
~~~~\left[\tilde{J}^i_m, \tilde{J}^j_n\right]
=\epsilon^{ijk}\tilde{J}^k_{m+n}-m\frac{\tilde{k}}{2}\delta^{ij}
\delta_{m+n,0},\nonumber\\
\alpha^{+i}_{ab}& = &
\frac{1}{2}\left(\delta^i_a\delta^0_b-\delta^i_b\delta^0_a\right)
+\frac{1}{2}\epsilon^{iab}, ~~~\tilde{c}=6\tilde{k}.
\label{small1}
\end{eqnarray}

In this algebra if we change $\tilde{G}^0_r$ to $-\tilde{G}^0_r$ then
the structure constants $\alpha ^{+i}_{ab}$ are changed to 
$\alpha^{-i}_{ab}$ where
\begin{equation}
\alpha^{-i}_{ab}=
-\frac{1}{2}\left(\delta^i_a\delta^0_b-\delta^i_b\delta^0_a\right)
+\frac{1}{2}\epsilon^{iab}.\label{alpham}
\end{equation}

Above (anti-) commutation relations can be derived from the following
OPEs of the small $N=4$ superconformal algebra.
\begin{eqnarray}
\tilde{T}(z)\tilde{T}(\omega)& = &\frac{\tilde{c}/2}{(z-\omega)^4}
+\frac{2\tilde{T}}{(z-\omega)^2}+
\frac{\partial\tilde{T}}{z-\omega},\nonumber\\
\tilde{T}(z)\tilde{G}^a(\omega)& = &
\frac{3/2\tilde{G}^a}{(z-\omega)^2}+
\frac{\partial\tilde{G}^a}{z-\omega},
~~\tilde{T}(z)\tilde{J}^i(\omega) =
\frac{\tilde{J}^i}{(z-\omega)^2}+
\frac{\partial\tilde{J}^i}{z-\omega},\nonumber\\
\tilde{G}^a(z)\tilde{G}^b(\omega)& = &
\frac{2\tilde{c}/3\delta^{ab}}{(z-\omega)^3}
-\frac{8\alpha^{+i}_{ab}\tilde{J}^i}{(z-\omega)^2}
+\frac{2\delta^{ab}\tilde{T}-4\alpha^{+i}_{ab}\partial 
\tilde{J}^i}{z-\omega},\nonumber\\
\tilde{J}^i(z)\tilde{G}^a(\omega)& = &\frac{\alpha^{+i}_{ab}
\tilde{G}^b}{z-\omega},
~~\tilde{J}^i(z)\tilde{J}^j(\omega) =
-\frac{\tilde{k}/2\delta^{ij}}{(z-\omega)^2}
+\frac{\epsilon^{ijk}\tilde{J}^k}{z-\omega}.
\label{small2}
\end{eqnarray}

A free field realizations of this algebra is given by the following 
expressions.
\begin{eqnarray}
\tilde{T}(z)& = &-J^0J^0+\frac{(k+1)}{\sqrt{k+2}}\partial J^0
-\frac{1}{k+2}J^iJ^i+\psi^a\partial\psi^a,\nonumber\\
\tilde{J}^1(z)& = &J^1-\psi^0\psi^1+\psi^2\psi^3, {\rm cyclic ~for }
~\tilde{J}^2 {\rm ~and} ~\tilde{J}^3,\nonumber\\
\tilde{G}^0(z)& = &2\left[J^0\psi^0+\frac{k+1}{\sqrt{k+2}}\partial\psi^0
+\frac{1}{\sqrt{k+2}}J^i\psi^i
+\frac{2}{\sqrt{k+2}}\psi^1\psi^2\psi^3\right],\nonumber\\
\tilde{G}^1(z)& = &2\left[J^0\psi^1+\frac{k+1}{\sqrt{k+2}}\partial\psi^1
+\frac{1}{\sqrt{k+2}}(-J^1\psi^0+J^2\psi^3-J^3\psi^2)\right.\nonumber\\
&{}&\left.-\frac{2}{\sqrt{k+2}}\psi^0\psi^2\psi^3\right],
\label{small3}
\end{eqnarray}
with cyclic expressions for $\tilde{G}^2$ and $\tilde{G}^3$. Here $k$ is
the level of the $SU(2)$ currents $J^i(z)$ used above. The
$SU(2)$ Kac-Moody algebra obeyed by them is 
\begin{equation}
J^i(z)J^j(\omega)=
-\frac{k/2\delta^{ij}}{(z-\omega)^2}
+\frac{\epsilon^{ijk}J^k(\omega)}{z-\omega}.\label{jij}
\end{equation}
Other two point functions are
\begin{equation}
\psi^a(z)\psi^b(\omega)=-\frac{1/2\delta^{ab}}{z-\omega},
~~~~~~~~~~J^0(z)J^0(\omega)=-\frac{1/2}{(z-\omega)^2}.
\label{2pt}
\end{equation}
The level $\tilde{k}$ occurring in the small algebra is related to  the
level $k$ of the ``free" $SU(2)$ by the equation $\tilde{k}=k+1$.
The central charge of this realization is $\tilde{c}=6\tilde{k}
=6(k+1)$.

We now summarize the large $N=4$ superconformal algebra, its 
OPEs and its most commonly used free field realization.
The corresponding symmetry is generated by sixteen operators. Apart 
from the energy-momentum tensor $T(z)$ and its four superpartners
$G^a(z)$ it has seven weight 1 currents $\{A^{\pm i}(z), U(z)\}$
generating $SU(2)\times SU(2)\times U(1)$ Kac-Moody subalgebra and
four weight 1/2 fermionic currents $Q^a(z)$. We shall denote the
corresponding modes by $\{L_m, G^a_r, A^{\pm i}_m, U_m, Q^a_r\}$
which satisfy the following large $N=4$ superconformal graded
Lie algebra.
\begin{eqnarray}
\left[L_m, L_n\right]& = &(m-n)L_{m+n}+\frac{c}{12}(m^3-m)
\delta_{m+n,0}\nonumber\\
\left[L_m, \Phi_n\right]& = &[(d_\Phi-1)m-n]\Phi_{m+n},\nonumber\\
&{}&\Phi_n\in\{G^a_n, A^{\pm i}_n, U_n, Q^a_n\}, 
d_\Phi\in\{3/2, 1, 1, 1/2\},\nonumber\\
\{G^a_r, G^b_s\}& = &
\frac{c}{3}(r^2-\frac{1}{4})\delta^{ab}\delta_{r+s,0}
+2\delta^{ab}L_{r+s}\nonumber\\
&{}&+4(r-s)[\gamma\alpha^{+i}_{ab}
A^{+i}_{r+s}+(1-\gamma)\alpha^{-i}_{ab}A^{-i}_{r+s}],
\nonumber\\
\left[A^{+i}_m, G^a_r\right]& = &
\alpha^{+i}_{ab}[G^b_{m+r}-2(1-\gamma)mQ^b_{m+r}],\nonumber\\
\left[A^{-i}_m, G^a_r\right]& = &
\alpha^{-i}_{ab}[G^b_{m+r}+2\gamma mQ^b_{m+r}],\nonumber\\
\left[A^{\pm i}_m, A^{\pm j}_n\right]& = &
\epsilon^{ijk}A^{\pm k}_{m+n}-
\frac{k^\pm}{2}m\delta^{ij}\delta_{m+n,0},\nonumber\\
\{Q^a_r, G^b_s\}& = &2(\alpha^{+i}_{ab}A^{+i}_{r+s}
-\alpha^{-i}_{ab}A^{-i}_{r+s})+\delta^{ab}U_{r+s},\nonumber\\
\left[A^{\pm i}_m, Q^a_r\right]& = &\alpha^{\pm i}_{ab}Q^b_{m+r},
~~\left[U_m, Q^a_r\right]=0, 
~~\left[U_m, G^a_r\right]=mQ^a_{m+r},\nonumber\\
\left[U_m, A^{\pm i}_n\right]& = &0,
~~~\{Q^a_r, Q^b_s\}=-\frac{c}{12\gamma(1-\gamma)}
\delta^{ab}\delta_{r+s,0},\nonumber\\
\left[U_m, U_n\right]& = &
-\frac{mc}{12\gamma(1-\gamma)}\delta_{m+n,0},~~
k^+=\frac{c}{6\gamma}, k^-=\frac{c}{6(1-\gamma)},\nonumber\\
\left[\alpha^{\pm i}, \alpha^{\pm j}\right]& = &
-\epsilon^{ijk}\alpha^{\pm k},\left[\alpha^{+i}, \alpha^{-j}\right]=0,
~~\{\alpha^{\pm i}, \alpha^{\pm j}\}=-\frac{1}{2}\delta^{ij},
\nonumber\\
c& = &\frac{6k^+k^-}{k^++k^-}, ~~~~~~~~\gamma=\frac{k^-}{k^++k^-}.
\label{large1}
\end{eqnarray}

This superalgebra has two independent
parameters unlike all the other superalgebras which have at the most
one parameter. In the present case these two parametres can be taken
either as $k^+$ and $k^-$ or as $c$ and $\gamma$. Above (anti-) 
commutators can be derived from the following OPEs.
\begin{eqnarray}
T(z)T(\omega)& = &\frac{c/2}{(z-\omega)^4}+\frac{2T}{(z-\omega)^2}
+\frac{\partial T}{z-\omega},\nonumber\\
T(z)\Phi(\omega)& = &\frac{d_\Phi\Phi}{(z-\omega)^2}
+\frac{\partial\Phi}{z-\omega},\nonumber\\
G^a(z)G^b(\omega)& = &\frac{2c/3\delta^{ab}}{(z-\omega)^3}
-\frac{8[\gamma\alpha^{+i}_{ab}A^{+i}+(1-\gamma)\alpha^{-i}_{ab}A^{-i}]}
{(z-\omega)^2}\nonumber\\
&{}&+\frac{\{2T\delta^{ab}
-4[\gamma\alpha^{+i}_{ab}\partial A^{+i}
+(1-\gamma)\alpha^{-i}_{ab}\partial A^{-i}]\}}
{z-\omega},\nonumber\\
A^{+i}(z)G^a(\omega)& = &\alpha^{+i}_{ab}[\frac{G^b}{z-\omega}
-\frac{2(1-\gamma)Q^b}{(z-\omega)^2}],\nonumber\\
A^{-i}(z)G^a(\omega)& = &\alpha^{-i}_{ab}[\frac{G^b}{z-\omega}
+\frac{2\gamma Q^b}{(z-\omega)^2}],\nonumber\\
A^{\pm i}(z)A^{\pm j}(\omega)& = &
-\frac{k^{\pm}/2\delta^{ij}}{(z-\omega)^2}
+\frac{\epsilon^{ijk}A^{\pm k}}{z-\omega},\nonumber\\
A^{+i}(z)A^{-j}(\omega)& = &0=A^{\pm i}(z)U(\omega),\nonumber\\
Q^a(z)G^b(\omega)& = &\frac{2[\alpha^{+i}_{ab}A^{+i}
-\alpha^{-i}_{ab}A^{-i}]+\delta^{ab}U}
{z-\omega},\nonumber\\
A^{\pm i}(z)Q^a(\omega)& = &
\frac{\alpha^{\pm i}_{ab}Q^b}{z-\omega},
~~~~U(z)Q^a(\omega)=0,\nonumber\\
U(z)G^a(\omega)& = &\frac{Q^a}{(z-\omega)^2},
~~~~~Q^a(z)Q^b(\omega)=-\frac{c}{12\gamma(1-\gamma)}
\frac{\delta^{ab}}{z-\omega},
\nonumber\\
U(z)U(\omega)& = &-\frac{c}{12\gamma(1-\gamma)}
\frac{1}{(z-\omega)^2}.
\label{large2}
\end{eqnarray}

This algebra attained the present form through the efforts of many people
\cite{adm}-\cite{iva}. Sevrin, Troost and van Proeyen 
gave the following free field realization also together with many
insights into its substructures.
\begin{eqnarray}
T(z)& = &-J^0J^0-\frac{1}{k+2}J^iJ^i+\psi^a\partial\psi^a,\nonumber\\
U(z)& = &\sqrt{k+2}J^0(z), ~~~~~Q^a(z)=\sqrt{k+2}\psi^a(z),\nonumber\\
A^{+1}(z)& = &J^1-\psi^0\psi^1+\psi^2\psi^3, {\rm cyclic ~for} ~A^{+2,3},
\nonumber\\
A^{-1}(z)& = &\psi^0\psi^1+\psi^2\psi^3, {\rm cyclic ~for}~A^{-2,3},
\nonumber\\
G^0(z)& = &2\left[J^0\psi^0+\frac{1}{\sqrt{k+2}}J^i\psi^i
+\frac{2}{\sqrt{k+2}}\psi^1\psi^2\psi^3\right],\nonumber\\
G^1(z)& = &2\left[J^0\psi^1
+\frac{1}{\sqrt{k+2}}(-J^1\psi^0+J^2\psi^3-J^3\psi^2)\right.\nonumber\\
&{}&\left.-\frac{2}{\sqrt{k+2}}\psi^0\psi^2\psi^3\right],
\label{large3}
\end{eqnarray}
with cyclic expressions for $G^2(z)$ and $G^3(z)$. Central
charge for this realization is $c=6(k+1)/(k+2)$ while other
parameters have values $k^-=1$, $k^+=k+1$, $\gamma=1/(k+2)$.
Above realizations is the most commonly used one in the literature.
This algebra too has been studied very extensively but We shall not
go into those aspects. Rather we now
turn to the middle $N=4$ superalgebra.

The middle $N=4$ superconformal algebra was discovered in \cite{ak}.
This algebra too has sixteen generators but the Kac-Moody subalgebra
this time is $SU(2)\times(U(1))^4$ generated by the weight 1 currents
$\hat{J}^i(z)$ and $\hat{U}^a(z)$ where $\hat{U}^a(z)\in\{\hat{U}(z)
=\hat{U}^0(z), \hat{U}^i(z)\}$. Here all $\hat{U}^a(z)$s are $U(1)$
currents. Other generators of this symmetry are denoted in this note by
$\hat{T}(z), \hat{G}^a(z)$ and $\hat{Q}^a(z)$ with conformal weights
2, 3/2 and 1/2 repectively. Modes of these operators 
are $\hat{J}^i_m, \hat{U}^a_m, \hat{L}_m, \hat{G}^a_r$
and $\hat{Q}^a_r$ respectively. These modes satisfy the following 
middle $N=4$ superconformal graded Lie algebra.
\begin{eqnarray}
\left[\hat{L}_m, \hat{L}_n\right]& = &(m-n)\hat{L}_{m+n}+
\frac{\hat{c}}{12}(m^3-m)\delta_{m+n,0},\nonumber\\
\left[\hat{L}_m, \hat{\Phi}_n\right]& = &
[(\hat{d}_\Phi-1)m-n]\hat{\Phi}_{m+n},\nonumber\\
&{}&
\hat{\Phi}_n\in\{\hat{G}^a_n, \hat{J}^i_n, \hat{U}^a_n, \hat{Q}^a_n\},
\hat{d}_\phi\in\{3/2, 1, 1, 1/2\},\nonumber\\
\{\hat{G}^a_r, \hat{G}^b_s\}& = &2\delta^{ab}\hat{L}_{r+s}
+4(r-s)\alpha^{+i}_{ab}\hat{J}^i_{r+s}
+\frac{\hat{c}}{3}(r^2-\frac{1}{4})\delta^{ab}\delta_{r+s,0},\nonumber\\
\left[\hat{J}^i_m, \hat{G}^a_r\right]& = &
\alpha^{+i}_{ab}\hat{G}^b_{m+r},
~~~[\hat{U}^i_m, \hat{G}^a_r]=2m\alpha^{-i}_{ab}\hat{Q}^b_{m+r},
\nonumber\\
\left[\hat{J}^i_m, \hat{J}^j_n\right]& = &\epsilon^{ijk}\hat{J}^k_{m+n}
-\frac{\hat{k}}{2}m\delta^{ij}\delta_{m+n,0},
~~~~~~~\left[\hat{U}^a_m, \hat{J}^i_n\right]=0,
\nonumber\\
\{\hat{Q}^a_r, \hat{G}^b_s\}& = &-2\alpha^{-i}_{ab}\hat{U}^i_{r+s}
+\delta^{ab}\hat{U}_{r+s},
~~~\left[\hat{U}^a_m, \hat{Q}^b_r\right]=0,
\nonumber\\
\left[\hat{J}^i_m, \hat{Q}^a_r\right]& = &
\alpha^{+i}_{ab}\hat{Q}^b_{m+r},
~~~\{\hat{Q}^a_r, \hat{Q}^b_s\}=
-\frac{\hat{k}}{2}\delta^{ab}\delta_{r+s,0},
\nonumber\\
\left[\hat{U}_m, \hat{G}^a_r\right]& = &m\hat{Q}^a_{m+r},~~~~
\left[\hat{U}^a_m, \hat{U}^b_n\right] =
-m\frac{\hat{k}}{2}\delta^{ab}\delta_{m+n,0}.
\label{middle1}
\end{eqnarray}

In reference \cite{ak} this algebra was obtained from 
the large algebra through a specific In\"on\"u-Wigner contraction.
Since this contraction procedure will be required again in the rest 
of this note, we begin by reviewing it.
First of all we make the following redefinitions
\begin{eqnarray}
{\hat{L}}_m& = &\lim_{\gamma\rightarrow 1}L_m,
~~~~~{\hat{U}}^i_m=\lim_{\gamma\rightarrow 1}\sqrt{1-\gamma}A^{-i}_m,
\nonumber\\
{\hat{G}}^a_r& = &\lim_{\gamma\rightarrow 1}G^a_r,
~~~~~{\hat{Q}}^a_r = \lim_{\gamma\rightarrow 1}\sqrt{1-\gamma}Q^a_r,
\nonumber\\
{\hat{J}}^i_m& = &\lim_{\gamma\rightarrow 1}A^{+i}_m,
~~~~~{\hat{U}}_m = \lim_{\gamma\rightarrow 1}\sqrt{1-\gamma}U_m.
\label{scaling1}
\end{eqnarray}
Some of these seem to have vanishing right hand sides and others
seem to be trivial. Neither of these observations is true because
above redefinitions are used to obtain the (anti-) commutators
(\ref{middle1}) from (\ref{large1}). This is done as follows. We
make above redefinitions in the (anti-) commutators of
the large algebra without taking the singular limit
$\gamma\rightarrow 1$ such that only hatted generators are used.
After that we take the limit $\gamma\rightarrow 1$. This limit 
makes some of the terms
vanish. In particular the currents $\hat{U}^i_m$ are abelianized.
Thus one of the $SU(2)$ Kac-Moody subalgebras of the large algebra
is changed to $(U(1))^3$. As a consequence the middle algebra gets
an $SU(2)\times(U(1))^4$ Kac-Moody subalgebra. We rename the level $k^+$
of the $SU(2)$ currents $A^{+i}_m$ as $\hat{k}$ for the currents
$\hat{J}^i_m$. Another effect of above singular rescalings is to reduce
the central charge to the value $\hat{c}=6\hat{k}$.

 It is worth noticing that due to inner $Z_2$ grading of the large
algebra an equivalent In\"on\"u-Wigner contraction would have been 
the following one.
\begin{eqnarray}
{\bar{L}}_m& = &\lim_{\gamma\rightarrow 0}L_m,
~~~~~{\bar{U}}^i_m=\lim_{\gamma\rightarrow 0}\sqrt{\gamma}A^{+i}_m,
\nonumber\\
{\bar{G}}^a_r&= &\lim_{\gamma\rightarrow 0}G^a_r,
~~~~~{\bar{Q}}^a_r = \lim_{\gamma\rightarrow 0}\sqrt{\gamma}Q^a_r,
\nonumber\\
{\bar{J}}^i_m& =&\lim_{\gamma\rightarrow 0}A^{-i}_m,
~~~~~{\bar{U}}_m = \lim_{\gamma\rightarrow 0}\sqrt{\gamma}U_m.
\label{scaling2}
\end{eqnarray}
This way we would have got
a middle algebra equivalent to (\ref{middle1}) with central charge
$\bar{c}=6\bar{k}=6k^-$ with roles of $\alpha^{+i}_{ab}$ and
$\alpha^{-i}_{ab}$ interchanged. Here $\bar{k}=k^-$ is the level of the 
$SU(2)$ algebra generated by the currents $\bar{J}^i_m$. Depending on
our convenience we shall use both of these contractions.

We have chosen the name {\it middle} for this new symmetry
since it has features common to both of the other symmetries.
Like the small symmetry it has only one $SU(2)$ subalgebra and does
not have $N=3$ as a subsymmetry and possibly can not be topologically
twisted. At the same time like the large algebra it has sixteen
generators out of which seven are of Kac-Moody nature. Unlike the 
small algebra it is {\it not} a subalgebra of the large algebra. 
Like the large algebra the middle one also has the small algebra as
a subalgebra.  More precisely the small algebra is an invariant
subalgebra of the middle algebra. Moreover the Kac-Moody subalgebra 
$SU(2)\times(U(1))^4$ is more stringent than $SU(2)$ but less than 
$SU(2)\times SU(2)\times U(1)$.

In reference \cite{ak} the OPEs
of the middle algebra were not derived. We shall do so now. 
We make following substitutions in eqs.(\ref{large2}) 
\begin{eqnarray}
\hat{T}(z)& = &\lim_{\gamma\rightarrow 1}T(z),
~~~~\hat{U}^i(z)=\lim_{\gamma\rightarrow 1}\sqrt{1-\gamma}A^{-i}(z),
\nonumber\\
\hat{G}^a(z)& = &\lim_{\gamma\rightarrow 1}G^{a}(z),
~~~~\hat{Q}^a(z)=\lim_{\gamma\rightarrow 1}\sqrt{1-\gamma}Q^{a}(z),
\nonumber\\
\hat{J}^i(z)& = &\lim_{\gamma\rightarrow 1}A^{+i}(z),
~~~~\hat{U}(z)=\lim_{\gamma\rightarrow 1}\sqrt{1-\gamma}U(z)
\label{scaling3}
\end{eqnarray}
together with the redefinitions $\hat{c}=\lim_{\gamma\rightarrow 1}c$
and $\hat{k}=\lim_{\gamma\rightarrow 1}k^+$. Just like earlier cases 
we take the limit $\gamma\rightarrow 1$ after making above 
redefinitions in the OPEs. As a result of this In\"on\"u-Wigner 
contraction many crucial changes
occur in equations containing the operators $\hat{U}^a$ and 
$\hat{Q}^a$. For example $\hat{U}^i$ terms disappear from 
$\hat{G}^a\hat{G}^b$ OPE. Other terms which disappear are 
$\hat{Q}^b$ in $\hat{J}^i\hat{G}^a$, $\hat{G}^b$ term
in $\hat{U}^i\hat{G}^a$, $\hat{U}^k$ term in $\hat{U}^i\hat{U}^j$,
$\hat{J}^i$ term in $\hat{Q}^a\hat{G}^b$ and
$\hat{Q}^b$ term in $\hat{U}^i\hat{Q}^a$. Most of the central
terms too are modified. These steps lead us 
to the following OPEs for the middle algebra.
\begin{eqnarray}
\hat{T}(z)\hat{T}(\omega)& = &\frac{\hat{c}/2}{(z-\omega)^4}
+\frac{2\hat{T}}{(z-\omega)^2}+
\frac{\partial\hat{T}}{z-\omega},\nonumber\\
\hat{T}(z)\hat{\Phi}(\omega)& = &\frac{\hat{d}_\Phi\hat{\Phi}}
{(z-\omega)^2}+\frac{\partial\hat{\Phi}}{z-\omega},
\nonumber\\
\hat{G}^a(z)\hat{G}^b(\omega)& = &
\frac{2\hat{c}/3\delta^{ab}}{(z-\omega)^3}
-\frac{8\alpha^{+i}_{ab}\hat{J}^i}{(z-\omega)^2}
+\frac{2\delta^{ab}\hat{T}-4\alpha^{+i}_{ab}\partial 
\hat{J}^i}{z-\omega},\nonumber\\
\hat{J}^i(z)\hat{G}^a(\omega)& = &\frac{\alpha^{+i}_{ab}
\hat{G}^b}{z-\omega},~~~~~~
\hat{U}^i(z)\hat{G}^a(\omega)=\frac{\alpha^{-i}_{ab}
\hat{Q}^b}{(z-\omega)^2}\nonumber\\
\hat{J}^i(z)\hat{J}^j(\omega)& = &
-\frac{\hat{k}/2\delta^{ij}}{(z-\omega)^2}
+\frac{\epsilon^{ijk}\hat{J}^k}{z-\omega},
~~~~\hat{U}^a(z)\hat{U}^b(\omega)=
-\frac{\hat{k}/2\delta^{ab}}{(z-\omega)^2},\nonumber\\
\hat{J}^i(z)\hat{U}^a(\omega)& = &0=\hat{U}^a(z)\hat{Q}^b,
~~~\hat{U}(z)\hat{G}^a(\omega)=\frac{\hat{Q}^a}{(z-\omega)^2},
\nonumber\\
\hat{Q}^a(z)\hat{G}^b(\omega)& = &\frac{\delta^{ab}\hat{U}-
2\alpha^{-i}_{ab}\hat{U}^i}{z-\omega},\nonumber\\
\hat{J}^i(z)\hat{Q}^a(\omega)& = &
\frac{\alpha^{+i}_{ab}\hat{Q}^b}{z-\omega}, 
~~~\hat{Q}^a(z)\hat{Q}^b(\omega)=-\frac{\hat{k}/2\delta^{ab}}
{z-\omega}.
\label{middle2}
\end{eqnarray}

It is a straightforward exercise to verify that these OPEs lead to
the middle superalgebra given in eqs.(\ref{middle1}). Once again
we should remember that by rescaling $A^{+i}(z), Q^a(z)$ 
and $U(z)$ instead
of $A^{-i}(z), Q^a(z)$ and $U(z)$ we would have got the OPEs of the
equivalent middle superconformal algebra.

If the middle algebra is part of the family of
$N=4$ superalgebras then, like the other two cases, symmetry 
corresponding to it too should have proper free field realizations. 
It turns out that there is an obstacle in achieving this goal.
To see it we start by In\"on\"u-Wigner
contraction of the free field realization (\ref{large3}) of the
large algebra because that is the easiest route for us owing to the
origin of the middle algebra. This time we shall use the other
rescalings given by eqs.(\ref{scaling2}) reason for which will become
apparent at the end of this process. Since for the realization
(\ref{large3}) we have $\gamma =1/(k+1)$
we shall do the following rescalings
\begin{eqnarray}
{\bar{T}}(z)& = &\lim_{k\rightarrow \infty}T(z),
~~~{\bar{U}}^i(z)=\lim_{k\rightarrow \infty}\frac{A^{+i}(z)}{\sqrt{k+1}},
\nonumber\\
{\bar{G}}^a(z)& = &\lim_{k\rightarrow \infty}G^a(z),
~~~{\bar{Q}}^a(z)=\lim_{k\rightarrow \infty}\frac{Q^a(z)}{\sqrt{k+1}},
\nonumber\\
{\bar{J}}^i(z) & = &\lim_{k\rightarrow \infty}A^{-i}(z),
~~~{\bar{U}}(z)=\lim_{k\rightarrow \infty}\frac{U(z)}{\sqrt{k+1}}
\label{scaling4}
\end{eqnarray}
for the operators occurring in the realization (\ref{large3}).
After making the redefinitions without taking the limit
we implement the contraction procedure by making $\gamma\rightarrow 0$ or
$k\rightarrow\infty$. This results in the following
free field realization of the middle algebra.
\begin{eqnarray}
\bar{T}(z)& = &-J^0J^0-U^iU^i+\psi^a\partial\psi^a,\nonumber\\
\bar{U}(z)& = &J^0(z), ~~~~~\bar{Q}^(z)=\psi^a(z),\nonumber\\
\bar{J}^1(z)& = &\psi^0\psi^1+\psi^2\psi^3, {\rm cyclic ~for}~\bar{J}^{2,3},
\nonumber\\
\bar{U}^i(z)& = &U^i(z), ~~\bar{G}^0=2(J^0\psi^0+U^i\psi^i),
\nonumber\\
\bar{G}^1(z)& = &2(J^0\psi^1-U^1\psi^0+U^2\psi^3-U^3\psi^2)
\label{middle3}
\end{eqnarray}
with cyclic expressions for ${\bar{G}}^2$ and ${\bar{G}}^3$.
Here $\bar{U}^a(z)$, $U^i(z)$ and $J^0(z)$ are all $U(1)$
currents. Particularly $U^i(z)$ is the abelianized set obtained by
contraction of $J^i(z)$. The central charge of this realization
is $\bar{c}=6k^-=6$. This is the problem with this realization
which has been alluded to earlier. It has a fixed value of the central 
charge. For effective use of a conformal symmetry its realizations 
with general central charges are desirable. Using above realization
the only way to obtain other central charges is to pile on several
copies, say $\bar{k}$,  of this realization to get $\bar{c}=6\bar{k}$.
This of course is not satisfactory because it involves a large
number of free fields while from experience we know that it is
not necessary to involve large number of fields to get a general
values of the central charge. 

It should be clarified that lack of a free field realization
with a general central charge does not cast any aspersions on the 
algebraic structure of this symmetry which conforms to all
the norms of a graded Lie superconformal symmetry. Problem lies exclusively
with the free field realizations. If this algebras does not have
proper free field realizations then it will end up just as a
mathematical curiosity with little applications to string theory
or 2D critical phenomena. This fear arises because of its
origin in In\"on\"u-Wigner contraction.  It is a well known
fact that In\"on\"u-Wigner contraction reduces the central charge
to either zero or a fixed value \cite{partha}.  Fortunately this will
not be so for the case of the middle algebra. We shall prove it below
by explicit construction but we can see formally also that this algebra
will have proper realizations. In usual contraction procedures the
central charge is used up for singular rescalings and is thus reduced
to a particular value, singular or otherwise. In such cases there is 
no other choice of rescalings
since central charge is the only free parameter in the algebra.
In the particular case under consideration we are lucky because the large
algebra has two free parameters, $c$ and $\gamma$ or $k^+$ and $k^-$,
as pointed out earlier. In our rescalings we use only one of them,
say $\gamma$, and therefore the central charge is still left as
a free parameter. This indicates that general free field realizations
of this algebra should exist.

To find proper realizations of the middle algebra we have to identify
the source of the problem in the existing realizations. To do
so we again start with the free field realization of the large
algebra. Let us consider
the $SU(2)$ currents in (\ref{large3}). We notice that the currents
$A^{+i}$ are realized in terms of another set $J^i$ of $SU(2)$ currents
supplemented by fermionic $SU(2)$ construction. On the other hand
currents $A^{-i}$ rely solely on fermionic $SU(2)$ construction and
do not have any genuine $SU(2)$ currents in them. In free field
realization (\ref{middle3}) of the middle algebra 
the currents $J^i$  have been abelianized leaving us with no
genuine $SU(2)$ currents a fact also confirmed by the expressions of
the currents $\bar{J}^i$. This is what leads us to a fixed central
charge since general values $6\bar{k}$ or $6\hat{k}$ come from
the level $\bar{k}$ or $\hat{k}$ of $SU(2)$ Kac-Moody algebra.
Moreover we can see that the second equivalent contraction
procedure for the realization (\ref{large3}) does not help either 
because in this case one has to take $\gamma\rightarrow 1$ which
demands $k=0$ giving $c=6(0+1)=6$ again. Thus we can solve 
the problem only if we can introduce a genuine $SU(2)$
set of currents in the realization of the middle $N=4$ symmetry.
This in turn requires that we do the contraction of such a
realization of the large algebra which has two genuine 
sets of ``free" $SU(2)$s.

In view of these circumstances we first take the problem of finding
a more general free field realization of the large algebra itself.
Above identification of the source of problem and a look at the
realization (\ref{large3}) immediately gives us an idea to proceed
further. We are tempted to introduce two sets $J^{\pm i}(z)$ of $SU(2)$
currents with levels $q^{\pm i}$ in the realization (\ref{large3})
such that $A^{\pm i}$s are defined as $J^{\pm i}$s supplemented 
by fermionic terms together
with suitable changes at other places in the realization. That
this naive attempt is inadequate can be seen as follows. Total central
charge contributed by the free fields in this case is
$3q^+/(q^++2)+3q^-/(q^-+2)+3$. Here first two terms are the
contributions of $J^{\pm i}$. 
Four fermions $\psi^a$ and the free boson $\phi(z)$ that gives
the $U(1)$ current $U(z)\propto J^0(z)=\partial\phi(z)$
contribute the last term $3=4(1/2)+1$. This central charge is,
in general, not the same as the required value which  at this point
is $6(q^++1)(q^-+1)/(q^++q^-+2)$ because now the currents 
$A^{\pm i}$ have levels $q^{\pm}+1$.

To overcome this new difficulty we shall not use generalised
Sugawara construction since we want a simple free field realization.
On the other hand we can not put a background charge on $\phi(z)$
since it will spoil the $U(1)$ nature of $J^0(z)$. What really has
happened is that introduction of three extra $SU(2)$ currents
has tilted the balance of free fields in favour of bosonic ones.
Again, in a realization, it is by no means necessary that numbers
of fermionic and
bosonic fields be equal but this is a good paradigm to stick 
to-particularly so for linear realizations of supersymmetry.
Once again we take up the expressions for $A^{\pm i}(z)$ and
introduce more fermions, say $\chi^a(z)$, in them parallel to
the existing ones. This step too leads us to similar problems.
Firstly, we have one fermionic free field more than the number of 
bosonic fields. Secondly, required and available central charges
still do not match. Free fields this time contribute a total of 
$3q^+/(q^++2)+3q^-/(q^-+2)+5$ while algebraic consistency demands
the value $6(q^++2)(q^-+2)/(q^++q^-+4)$; currents 
$A^{\pm i}$ now having levels $q^\pm+2$. The first problem is solved by
introducting an extra free boson $\varphi(z)$ in the realization
which increases the contribution of free fields to the central charge to 
$3q^+/(q^++2)+3q^-/(q^-+2)+6$. This, of course, is still  
deficient by an amount
\begin{eqnarray}
\Delta c& = &\frac{6(q^++2)(q^-+2)}{q^++q^-+4}
-\frac{3q^+}{q^++2}-\frac{3q^-}{q^-+2}-6\nonumber\\
& = &6\left(\frac{c-6}{\sqrt{6c}}\right)^2
\equiv 6\beta_0^2. 
\label{deltac}
\end{eqnarray}
But this deficient central charge can now be compensated by putting 
suitable background Coulomb charges on free bosons.
Earlier restriction on background charge is no more a problem.
Since there are two bosons now they will always have a
linear $U(1)$ combination. We shall
get above deficient central charge if we put background charges
$a\beta_0$ and $b\beta_0$ on $\phi$ and $\varphi$ repectively
such that $a^2+b^2=1$. Since $\gamma$ decides relative weightage
of inner $Z_2$ grading in this algebra the easiest guess for $a$ 
and $b$ are $a=\sqrt{\gamma}$ and $b=-\sqrt{1-\gamma}$. Reason
for an extra negative sign in $b$ will become clear shortly.

Above arguments fix the form of $T(z)$ completely which is
\begin{eqnarray}
T(z)& = &-J^0J^0+\sqrt{\frac{\gamma}{6c}}(6-c)\partial J^0
-\frac{6\gamma}{c}J^{+i}J^{+i}+\psi^a\partial\psi^a
+\chi^a\partial\chi^a
\nonumber\\
&{}&-K^0K^0-\sqrt{\frac{1-\gamma}{6c}}(6-c)\partial K^0
-\frac{6(1-\gamma)}{c}J^{-i}J^{-i},
\label{t4}
\end{eqnarray}
where $K^0=\partial\varphi$.
We can now very easily get the expression for the $U(1)$ current
$U(z)$. It has to be a linear combination of $J^0(z)$ and
$K^0(z)$. Requirements of correct OPEs with 
$T(z)$ and with itself completly fix the linear combination
which is
\begin{equation}
U(z)= \sqrt{\frac{c}{6\gamma}}J^0(z)
+\sqrt{\frac{c}{6(1-\gamma)}}K^0(z).
\label{u4}
\end{equation}
If we had chosen $b$ above with a positive sign then sign of $K^0$
in (\ref{u4}) would have also changed.
Expressions for the two $SU(2)$ currents also are very simple
to write. These are generalizations of the expressions in 
(\ref{large3}) to incorporate $\chi^a$ and $J^{-i}$. Thus
we get
\begin{eqnarray}
A^{+1}(z)& = &
J^{+1}-\psi^0\psi^1+\psi^2\psi^3-\chi^0\chi^1+\chi^2\chi^3,
{\rm cyclic ~for}~A^{+2,3},\nonumber\\
A^{-1}(z)& = &
J^{-1}+\psi^0\psi^1+\psi^2\psi^3+\chi^0\chi^1+\chi^2\chi^3,
{\rm cyclic ~for}~A^{-2,3}.\label{apm4}
\end{eqnarray}

Expressions for $G^a(z)$s too are easy to write but they are
tedious to verify.
They will have $\{J^0, J^{+i}, \psi^a\}$ terms as in the realization
(\ref{large3}) and similar terms in  $\{K^0, J^{-i}, \chi^a\}$. 
Since OPEs $G^a(z)G^a(\omega)$ (no sum on $a$) give the energy-momentum
tensor which now has background terms, we must make 
provision for generating such terms. We do so by adding partial 
derivatives of $\psi^a$ and $\chi^a$ to the expressions of the 
supercurrents $G^a$. Coefficients of these partial derivatives
are fixed by the requirements that correct background charge terms of
$T(z)$ are generated in above mentioned OPEs. The coefficients of 
$J^{+ i}\psi$ and $J^{-i}\chi^a$ type terms in $G^a(z)$ are fixed 
by the requirement that $G^aG^b$, where $a\neq b,$
OPEs give correct coefficients of the $SU(2)$ currents on the right hand
side. Explicit calculations give the following forms for the
supercurrents
\begin{eqnarray}
G^0(z)& = &2[J^0\psi^0+\sqrt{\frac{\gamma}{6c}}(6-c)\partial\psi^0
+K^0\chi^0-\sqrt{\frac{(1-\gamma)}{6c}}(6-c)\partial\chi^0
\nonumber\\
&{}& +\sqrt{\frac{6\gamma}{c}}(J^{+i}\psi^i+2\psi^1\psi^2\psi^3)
+\sqrt{\frac{6(1-\gamma)}{c}}(J^{-i}\chi^i+2\chi^1\chi^2\chi^3)],
\nonumber\\
G^1(z)& = &2[J^0\psi^1+\sqrt{\frac{\gamma}{6c}}(6-c)\partial\psi^1
+K^0\chi^1-\sqrt{\frac{(1-\gamma)}{6c}}(6-c)\partial\chi^1\nonumber\\
&{}&+\sqrt{\frac{6\gamma}{c}}(-J^{+1}\psi^0+J^{+2}\psi^3-J^{+3}\psi^2
-2\psi^0\psi^2\psi^3)\nonumber\\
&{}&+\sqrt{\frac{6(1-\gamma)}{c}}(-J^{-1}\chi^0+J^{-2}\chi^3-J^{-3}\chi^2
-2\chi^0\chi^2\chi^3)]\label{ga4}
\end{eqnarray}
with cyclic expressions for $G^2$ and $G^3$.
Finally the expressions for $Q^a(z)$s are obtained by calculating
the OPEs $U(z)G^a(\omega)$ and comparing them with the required
form. This way we get the following expressions
\begin{eqnarray}
Q^a(z)& = &\sqrt{\frac{c}{6\gamma}}\psi^a(z)
+\sqrt{\frac{c}{6(1-\gamma)}}\chi^a(z).\label{qa4}
\end{eqnarray}

We have now all the operators of the large $N=4$ superconformal symmetry
in terms of sixteen ``free" fields $J^{\pm i}, \phi, \varphi,
\psi^a$ and $\chi^a$. Expressions (\ref{t4})-(\ref{qa4})
give a realization of the large $N=4$ 
superconformal symmetry which is complete
in all aspects relevant for us. It has two genuine $SU(2)$ 
currents $J^{\pm i}$ with
levels $q^\pm$ which lead to proper realizations of $A^{\pm i}$ with
levels $k^+=q^++2=c/(6\gamma)$ and $k^-=q^-+2=c/(6(1-\gamma))$. In
it the number of free field is equal to the number of generators of the
symmetry. Finally we would like to remark that these 
expressions should be closely related to the 
generators of the $N=4$ supersymmetry of a non-linear 
sigma model studied in \cite{iva} that have remained largely unnoticed.

Above realization can immediately be used to obtain new free field
realizations of the small $N=4$ and $N=3$ superconformal symmetries
since both of them are subsymmetries of the large one.
One also notices that in this realization the Coulomb charge terms 
vanish precisely for $c=6$. This observation explains the special 
nature of realizations with $c=6$ for $N=4$ superconformal algebras 
and their common occurrence. In above realization one can ignore the
$SU(2)$ currents and the algebra is realized with the help of ten
free fields including eight fermions and two bosons. 

At this point we have all the ingredients required to obtain 
a complete free field realization of the middle algebra. 
As expected realization (\ref{t4})-(\ref{qa4}) can be contracted in two
ways to give two equivalent realizations of the middle $N=4$
superconformal symmetry.  By doing singular rescalings of the operators 
$A^{-i}, J^{-i}, Q^a,$ and $U$ we get the following expressions for
the operators of the middle $N=4$ superconformal symmetry.
\begin{eqnarray}
\hat{T}(z)& = &-J^0J^0+\frac{6-c}{\sqrt{6c}}\partial J^0
-K^0K^0-\frac{6}{c}J^{+i}J^{+i}-\frac{6}{c}V^iV^i
\nonumber\\
& &+\psi^a\partial\psi^a +\chi^a\partial\chi^a,\nonumber\\
\hat{U}(z)& = &\sqrt{\frac{c}{6}}K^0(z), ~~~~~\hat{Q}^a(z)=
\sqrt{\frac{c}{6}}\chi^a(z),~~\hat{U}^i(z)=V^i(z),\nonumber\\
\hat{J}^1(z)& = &J^{+1}-\psi^0\psi^1+\psi^2\psi^3
-\chi^0\chi^1+\chi^2\chi^3,{\rm cyclic ~for}~\hat{J}^{2,3},
\nonumber\\
\hat{G}^0& = &2\{J^0\psi^0+\frac{6-c}{\sqrt{6c}}\partial\psi^0
+K^0\chi^0\nonumber\\
& &+\sqrt{\frac{6}{c}}(J^{+i}\psi^i+2\psi^1\psi^2\psi^3)
+\sqrt{\frac{6}{c}}V^i\psi^i\},\nonumber\\
\hat{G}^1(z)& = &2\{J^0\psi^1
+\frac{6-c}{\sqrt{6c}}\partial\psi^1+K^0\chi^1
\nonumber\\
&{}&+\sqrt{\frac{6}{c}}(-J^{+1}\psi^0+J^{+2}\psi^3-J^{+3}\psi^2
-2\psi^0\psi^2\psi^3)\nonumber\\
&{}&+\sqrt{\frac{6}{c}}(-V^1\psi^0+V^2\psi^3-V^3\psi^2)\}.
\label{middle4}
\end{eqnarray}
with cyclic expressions for $\hat{G}^2(z)$ and $\hat{G}^3(z)$.
Here the operators $V^i(z),$ ${\hat{U}}(z)$and ${\hat{U}}^i(z)$
are all $U(1)$ currents; $V^i(z)$s being the abelianized version of 
$J^{-i}(z)$s. This realization of the middle $N=4$
superconformal symmetry, having a central charge
$\hat{c}=6\hat{k}=6(q^++2)$,
is complete in all aspects discussed above. Leaving out the
generators ${\hat{U}}(z)$, ${\hat{U}}^i(z)$ and ${\hat{Q}}^a(z)$
from eqs.(\ref{middle4}) we get another realization for the small symmetry.

\begin{center}
 {\it Table A : Classification of Superconformal Symmetries}

\begin{tabular}{|c|c|c|c|c|}
\hline
S.N. &  Name and      & No. of     & List of    & Kac-Moody\\
{}   & Central Charge & Generators & Generators & Subalgebra\\
\hline
1.   &$N=0, c $       & One        & $T(z)$     & -        \\
\hline
2.   &$N=1, c  $      &Two         &$T(z), G(z)$& -        \\
\hline
3.   &$N=2, c  $      &Four        &$T(z), G^\pm(z),$&$U(1)$\\
{}   &${}$      &{}        &$U(z)$&{}\\
\hline
4.  &$N=3$      &Eight&$T(z), G^i(z),$&$SU(2)$\\
{}  &$c=3k/2$   &{}   &$J^i(z), Q(z)$&{}\\
\hline
{} &{}&{}&{}&{}\\
5.&Small $N=4$&Eight&$\tilde{T}(z),
\tilde{G}^a(z),$&$SU(2)$\\
{}&$\tilde{c}=6\tilde{k}$&{}&$\tilde{J}^i(z)$&${}$\\
\hline
{} &{}&{}&{}&{}\\
{}&Middle $N=4$&Sixteen&$\hat{T}(z),\hat{G}^a(z),$&$SU(2)\times$\\
6.&$\hat{c}=6\hat{k}$&{}&$\hat{J}^i(z), \hat{U}^a(z),$&
$(U(1))^4$\\
{}&{}&{}&$\hat{Q}^a(z)$&{}\\
\hline
{} &{}&{}&{}&{}\\
{}&Large $N=4$&Sixteen&$T(z),G^a(z),$&$SU(2)\times$\\
7.&$c=\frac{6k^+k^-}{(k^++k^-)}$&{}&$A^{\pm i}(z), U(z),$&$SU(2)\times$\\
{}&${}$&{}&$Q^a(z)$&$U(1)$\\
\hline
\end{tabular}
\end{center}

In view of the above results we conclude that the middle algebra has
(anti-) commutators, OPEs and a free field realization on par with the
other superconformal symmetries and should be treated as such. The
complete classification of $N=4$ superconformal algebras therefore has
three members-large, middle and small. This classification is complete
in the sense that we can not further contract the middle algebra or
the small algebra to obtain more $N=4$ superalgebras because in doing
so we shall lose the remaining parameter, the central charge. This 
classification of $N=4$ together with other superconformal algebras is 
summarized in Table A.

To conclude we have re-examined the classification of $N=4$
superconformal symmetries. We have proposed to enhance the number 
of such symmetries from two to three by studying some essential 
aspects of a new $N=4$ symmetry called middle one in this note.
Since large and small algebras have been very extensively studied in
literature the same can be done for the middle algebra too because 
we now have its OPEs and a complete free field realization. Moreover
the realization (\ref{t4})-(\ref{qa4}) itself should lead to renewed 
inestigations of many aspects of the large algebra. 

Results of this note will have numerous applications. These 
are broadly divided into three
categories. In the first category we include the study of the
various aspects of middle and large symmetries themselves 
while the second one pertains to uses of above free field realizations
and their decendents. The third category is related to occurrences
of the middle $N=4$ symmetry in string theory.

In the light of above realization (\ref{t4})-(\ref{qa4})
we can generalize the existing Kazama-Suzuki type constructions 
\cite{vst,gk} for the large $N=4$ superconformal algebra to 
incorporate real $SU(2)$s in both sets of the $A^{\pm i}(z)$s.
We expect that more general constructions rather than simple Wolf 
spaces \cite{wolf} are relevant for the large algebra. This in turn
can be used to obtain Kazama-Suzuki realizations of the middle
symmetry. One could also ask whether there is a nonlinear algebra 
corresponding to the middle algebra \cite{nonlinear}. Above free field 
realizations should also be derivable by Hamiltonian reduction \cite{ito}. 
Another open problem is the study of corresponding nonlinear sigma models
\cite{rocek} which also will be useful to study 
superstring actions with $N=4$ world sheet supersymmetry
combined with manifest target space duality \cite{gpr}
along the lines of \cite{aa1} or \cite{fawad}.

Above free field realization of the large symmetry will be quite
useful in the applications of $N=2$ topological superconformal field 
theories \cite{top2}.
Screening operators and related aspects of the middle algebra on the lines 
of ref.\cite{screen} can also be investigated. 

Few years ago it was observed that $N$ strings can be obtained as 
a special class of vacua of $N+1$ strings \cite{vafa}. It will be
interesting to find out the place of the {\it middle} $N=4$ string 
in this hierarchy. 
A route to find the critical dimensions of the corresponding string
will be to do the BRST quantization of a system with middle algebra
as constraints on the lines of ref.\cite{ck}. We also suspect that
this symmetry could be relevant for the Coulomb and Higgs branches 
of (4, 4) supersymmetric field theories in two dimensions\cite{ds}
as well as for strings moving on solitons \cite{kpr,chs} which have
at least the small algebra. 
In the AdS-CFT correspondence the relevant space time symmetry
for superstring propagating on curved spacetime manifold
${\cal M}= AdS_3\times S^3\times T^4$ should be the middle $N=4$
superconformal symmetry itself \cite{adscft3, hw, aa2}. 
We expect to return to some these problems in near future.

{\it Acknowledgements.} This work was started at the Institute of
Mathematical Sciences, Chennai as a postdoctoral fellow. Part of
the calculations were done during a visit to the Abdus Salam ICTP,
Trieste for which I am grateful to S. Randjbar-Daemi.
I thank Alok Kumar for introducing me to the subject, K.S. Narain
for discussions and H.S. Mani for hospitality at the Mehta
Research Institute, Allahabad. Also I thank all of them as well as 
Hashim Rizvi and S.K. Singh for encouragement. D.P. Jatkar read the 
manuscript and made useful suggestions for which I am thankful
to him.


\begin{thebibliography}{xxx}
\bibitem{mcena} J. Maldacena, Adv. Theor. Math. Phys. 2(1998)231
\bibitem{adscft1} S. Gubser, I. Klebanov and A. Polyakov,
Phys. Lett. B428(1998)105
\bibitem{adscft2} E. Witten, Adv. Theor. Math. Phys. 2(1998)253
\bibitem{adscft3} A. Giveon, D. Kutasov and N. Seiberg,
Adv. Theor. Math. Phys. 2(1998)733; S. Elitzur, O. Feinerman,
A. Giveon and D. Tsabar, Phys. Lett. B449(1999)180;
I. Pesando, JHEP 02(1999)007
\bibitem{adscft4} K. Ito, Phys. Lett. B449(1999)48;
O. Andreev, hep-th/9901118;
J. de Boer, A. Pasquinucci and K. Skenderis, hep-th/9904073;
J. Rasmussen, hep-th/9905181
\bibitem{ak0} A. Ali and A. Kumar, Mod. Phys. Lett. 
A8(1993)2045; G. Horowitz and D. Welch, Phys. Rev. Lett.
71(1993)328; N. Kaloper, Phys. Rev. D48(1993)2598
\bibitem{hw} S.F. Hassan and S.R. Wadia, Phys. Lett. B402(1997)43;
Nucl. Phys. B526(1998)311
\bibitem{sw} N. Seiberg and E. Witten, hep-th/9903224
\bibitem{adm} M. Ademollo, L. Brink, A. d'Adda, R. d'Auria, 
E. Napolitano, S. Sciuto, E. del Giudice, P. Di Vecchia, S. Ferrara,
F. Gliozzi, R. Musto and R. Pettorino, Phys. Lett. B62(1976)105;
Nucl. Phys. B114(1976)297
\bibitem{stv} A. Sevrin, W. Troost and A. van Proeyen, Phys. Lett.
B208(1988)447; Ph. Spindel, A.Sevrin, W.Troost and A. van Proeyen,
Nucl. Phys. B311(1988/89)465
\bibitem{sch1} K. Schoutens, Phys. Lett. B194(1987)75; Nucl. Phys.
B295(1988)634
\bibitem{sch2} A. Schwimmer and N. Seiberg, Phys. Lett. B184(1987)191
\bibitem{iva} E. Ivanov, S. Krivonos and V. Leviant, Nucl. Phys.
B304(1988)601; Phys. Lett. B215(1988)689
\bibitem{ak} A. Ali and A. Kumar, Mod. Phys. Lett. A8(1993)1527
\bibitem{partha} P. Majumdar, J. Math. Phys. 34(1993)2059
\bibitem{vst} A. van Proeyen, Class. Quant. Grav. 6(1989)1501;
A. Sevrin and G. Theodoridis, Nucl. Phys. B332(1990)380
\bibitem{gk}S.J. Gates, Jr. and S.V. Ketov, Phys. Rev. D52(1995)2278
\bibitem{wolf} J.A. Wolf, J. Math. Mech. 14(1965)1033; D.V. 
Alekseevskii, Math. USSR Izv. 9(1975)297
\bibitem{nonlinear} P. Goddard and A. Schwimmer, Phys. Lett.
B214(1988)209
\bibitem{ito} K. Ito, J.O. Madsen and J.L. Petersen, Phys. Lett.
B318(1993)315
\bibitem{rocek} M. R\^ocek, K. Schoutens and A. Sevrin, Phys. Lett.
B265(1991)303; M. G\"unaydin, Phys. Rev. D47(1993)3600; S.J. Gates, Jr.
and L. Rana, Phys. Lett. B345(1995)233
\bibitem{gpr} A. Giveon, M. Porrati and E. Rabinovici, Phys. Rep.
244(1994)71
\bibitem{aa1} A. Ali, Phys. Rev. D52(1995)2312
\bibitem{fawad} S.F. Hassan, Nucl. Phys. B460(1996)362
\bibitem{top2} A. Ali, D.P. Jatkar and A. Kumar, preprint
IP/BBSR/91-14; A. Ali, A. Kumar, J. Maharana and G. Sengupta,
Commun. Math. Phys. 148(1992)117; S. Nojiri, Phys. Lett. B264(1992)264
\bibitem{screen} M. G\"unaydin, J.L. Petersen, A. Taormina and 
A. van Proeyen, Nucl. Phys. B322(1989)402; K. Ito, J.O. Madsen and
J.L. Petersen, Phys. Lett. B292(1992)298
\bibitem{vafa} N. Berkovits and C. Vafa, Mod. Phys. Lett. A9(1994)653;
F. Bastianelli, N. Ohta and J.L. Petersen, Phys. Lett. B327(1994)35;
Phys. Rev. Lett. 73(1994)1199; N. Ohta and T. Shimizu, Phys. Lett.
B355(1995)127; S.V. Ketov, Class. Quant. Grav. 12(1995)925;
S.J. Gates, Jr., Phys. Lett. B338(1994)31
\bibitem{ck} D. Chang and A. Kumar, Phys. Rev. D35(1987)1388;
F. Bastianelli and N. Ohta, Phys. Rev. D50(1994)4051
\bibitem{ds} D.-E. Diaconescu and N. Seiberg, JHEP 07(1997)001
\bibitem{kpr} C. Kounnas, M. Porrati and B. Rostand, Phys. Lett. 
B258(1991)61
\bibitem{chs} C. Callan, J. Harvey and A. Strominger,
Nucl. Phys. B359(1991)611; Nucl. Phys. B367(1991)60
\bibitem{aa2} A. Ali, work in progress.
\end{thebibliography}
\end{document}